# Long- and Short-Ranged Chiral Interactions in DNA-Assembled Plasmonic Chains

*Kevin Martens[1], Felix Binkowski[2], Linh Nguyen[1], Li Hu[3], Alexander O. Govorov[4], Sven Burger[2, 5] and Tim Liedl[1\*]*

[1]*Faculty of Physics, Ludwig-Maximilians-University, Geschwister-Scholl-Platz 1, 80539 Munich, Germany*
[2]*Zuse Institute Berlin, Takustraße 7, D-14195 Berlin, Germany*
[3]*Chongqing Engineering Laboratory for Detection, Control and Integrated System, Chongqing Technology and Business University, Chongqing 400067, China*
[4]*Department of Physics and Astronomy, Nanoscale and Quantum Phenomena Institute, Ohio University, Athens, Ohio 45701, United States*
[5]*JCMwave GmbH, Bolivarallee 22, 14050 Berlin, Germany*
\* *tim.liedl@lmu.de*

**Molecular chirality plays a crucial role in innumerable biological processes. The chirality of a molecule can typically be identified by its characteristic optical response, the circular dichroism (CD). CD signals have thus long been used to identify the state of molecules or to follow dynamic protein configurations. In recent years, the focus has moved towards plasmonic nanostructures, as they show potential for applications ranging from pathogen sensing to novel optical materials. The plasmonic coupling of the individual elements of such chiral metallic structures is a crucial prerequisite to obtain sizeable CD signals. We here identified and implemented various coupling entities – chiral and achiral – to obtain chiral transfer over distances close to 100 nm. The coupling is realized by an achiral nanosphere situated between a pair of gold nanorods that are arranged far apart but in a chiral fashion. We synthesized these structures with nanometer precision using DNA origami and obtained sample homogeneity that allowed us to directly demonstrate efficient chiral energy transfer between the distant nanorods. The transmitter particle causes a strong enhancement in amplitude of the CD response, the emergence of an additional chiral feature at the resonance frequency of the nanosphere, and a redshift of the longitudinal plasmonic resonance frequency of the nanorods. Numerical simulations closely match our experimental observations and give insights in the intricate behavior of chiral optical fields and the transfer of plasmons in complex architectures.**

Chirality describes a geometric feature of structures that do not have any internal planar symmetry. As a consequence, a chiral structure and its mirror image cannot be brought to coincide with each other through the geometrical transformations of rotation and translocation. Such objects of opposite handedness are called enantiomers. They play a decisive role in nature, as a right-handed molecule can have significantly different functions in biological systems as its left-handed counterpart.[1] Along with their distinctive geometrical and thus chemical features, chiral molecules exhibit intricate optical responses upon irradiation with linearly and circularly polarized light. These phenomena are called optical rotatory dispersion (ORD) and circular dichroism (CD), respectively. Amongst others, CD allows to monitor folding processes of proteins[2] and to evaluate the chiral quality of synthetic chemicals.[3] Next to molecular identity and function, chirality enables the creation of designer architectures such as chiral photonic crystals[4,5] and chiral metamaterials.[6-8] Optical activity can already arise from chiral nanoparticles[9-12] or assemblies of several achiral NPs into chiral structures through plasmon-plasmon interactions between the surfaces of the NPs.[13-18] CD responses of chiral plasmonic systems show great potential in applications ranging from chiral discrimination of molecules[19] and sensing,[20-22] over enantiomer-selective catalysis[23] to circular polarizing devices.[7]

Next to CD signals arising from pure organic compounds on the one side and inorganic particles or assemblies on the other side also interactions between these two domains can occur and give rise to strong effects. Particularly, plasmonic surfaces and particles can strongly increase the CD signals of chiral biomolecules in their vicinity, which in turn can enhance the sensitivity of chiroptical detection of biomolecules.[24-30] In such experiments, the strong, plasmon-induced electromagnetic (EM) near-field couples to the chiral near-field of the biomolecules with the latter having its maximum in the UV and extending only a weak tail into the visible spectral range. This type of "CD transfer" leads to augmentation of the signal strength in the plasmonic window that predominantly occurs in the visible and near infrared (NIR).

DNA nanotechnology[31] and in particular DNA origami[32,33] proved itself to be a powerful tool to implement complex plasmonic particle assemblies with nanometer accuracy.[34] DNA origami structures are formed from a long single-stranded DNA, serving as a "scaffold" strand, folded into shape by hundreds of synthetic "staple" oligonucleotides.[31] NPs can be attached by functionalizing them with thiol-modified oligonucleotides that then hybridize at specific positions on the origami structure. These features make DNA origami an ideal platform for nanostructures with tailored optical functionalities.[16-18]

DNA origami has been used to achieve chiral nanorod (NR) assemblies in a variety of ways, where rods crossing each other in an X-shape or an L-shape are predominant.[35,36] Switchable variants of these geometries[37-40] further allow for sensitive detection of biomolecules.[22] In all of these assemblies, the distances between the nanoparticles play a crucial role, since plasmon-plasmon interactions are usually limited to the range of a few nanometer.[35,36,38-40] Via sufficiently small gaps, plasmonic energy can traverse efficiently over NP chains[41] and plasmon transfer can even occur between NPs with non-identical resonance frequencies by quasi-occupation of different transfer channels.[42] Transfer of chiral signatures over chains of particles has been predicted theoretically by authors of this study,[43] however, experimental prove has been lacking.

Here, we explore this new type of transfer over long distances. In our experiments, plasmon-assisted chiral interactions occur in chiral assemblies of two nanorods arranged with a surface-to-surface distance of over 60 nm. The presence of a third, spherical transmitter particle in the gap between the rods efficiently couples the near-field of the rods leading to strong signal increase of the longitudinal modes and the evolution of new CD features in the spectral range of the spherical particle.

We designed and synthesized a compound DNA origami platform composed of two individual structures (Fig. 1; see Supplementary Information note S1 for design details). The full structure has an overall length of 100 nm and, by using thiol-DNA functionalization and specific handle strands, accommodates two gold NRs (each 54 nm long and 23 nm wide) at its ends. The rods are designed to have a surface-to-surface distance of 62 nm and they are tilted by 90° in respect to each other when observed along the axis of the origami structure (Fig. 1b). Also in this perspective, each one of the NRs overlap, resulting in an L-shaped object. To serve as a plasmonic transmitter, a 40 nm gold nanosphere (NS) can be attached in between the NRs, spaced 12 nm from each NR (Fig. 1a).

To investigate the effect of this transmitter particle, we synthesized samples with only NRs (NR– –NR) and samples with the center nanosphere present (NR–NS–NR). Transmission electron microscopy (TEM) confirmed the assemblies as designed (Fig. 2 a, b). Note that the angular correlation between the particles is lost in the TEM micrographs, as the DNA origami structures – that were assembled in solution and will also perform their task in solution – needed to adsorb and dry on the TEM grids before being imaged in vacuum.

A study of 200 assemblies of the sample NR– –NR yielded a majority (58 %) to be in the expected arrangement (Fig. 2c, Supplementary Note S3). The most common disarrangement was an extra nanorod attached to the origami structure (NR– –2NR; 15 %), which occurs when 2 NRs attach to the handles designed for a single NR.

Another 14 % of the structures contained a NS instead of a NR on one side (NS– –NR) due to spurious NSs remaining from NR synthesis. On the other hand, a study of 300 assemblies of the sample NR–NS–NR resulted in 51 % well-assembled structures. The most common disarrangements here were structures with a NS instead of a NR attached to one of the ends (NS–NS–NR), contributing to 18 % of the sample, as well as assemblies lacking the spherical particle (NR– –NR), contributing to 9 % of all assemblies.

The CD measurements were performed at concentrations of ~ 0.1 nM of the plasmonic assemblies as estimated by the extinction of the NRs longitudinal mode. To compensate for varying sample concentrations, the CD signals were normalized by the same maximum extinction amplitude for each sample. As shown in Fig. 3, the CD spectrum of the NR– –NR sample shows a typical CD signal for right-handed chiral L-shaped nanostructures with a maximum dip at 648 nm and a peak at 698 nm around the NRs longitudinal plasmon resonance frequency of 676 nm (Fig. 3a, b). In comparison, the NR–NS–NR sample shows the same dip-peak signature but with a 3.5-fold increased amplitude. The peaks are red-shifted to 657 nm (minimum) and 704 nm (maximum), matching a shift in the extinction spectrum for the longitudinal NR peak to a wavelength of 681 nm (Fig 3b). Additionally, a new CD signature appears in the NR–NS–NR sample around the NS resonance wavelength. At around 512 nm a slight positive deflection can be observed followed by a pronounced negative deflection at 557 nm, with the signal crossing the zero-line close to the plasmonic resonance frequency of the NS at ~530 nm (Fig. 3a, inset).

We computed CD spectra of the various nanoparticle arrangements by numerically solving the time-harmonic, linear Maxwell's equations. Single NR– –NR and NR–NS–NR arrangements are enclosed in the computational domain. We use tabulated material data for NP material Au[44] and a constant refractive index of 1.4 for the background material. Circularly polarized plane waves of various wavelengths and incidence directions are used as excitation. From the near-field solutions, CD spectra and extinction spectra are obtained (see Supplementary Note S5). The simulated spectra are shown in Fig. 3c and 3d next to heat maps visualizing the coupling via the near-field around the particles (Fig. 3e and 3f). Strikingly, our models reproduce all the characteristic features of the experimental results.

When compared to the simulations, the recorded data shows many consistencies as well as some noticeable differences. The main and obvious discrepancy between theory and experiment are the broadened dips and peaks and the less pronounced enhancement factor in the experimental data. These differences can be accounted for by the inhomogeneity of the samples, as only small fractions of aggregates can manifest themselves in spectral broadening, which in turn is accompanied by reduced signal increase in case of the transmitter particle present. Notably, due to their achiral nature, the largest part of the disarrangements described above (NS–NS–NR and NS– –NR in the NR–NS–NR sample) cannot be a source of the increased signal strength observed for the NR–NS–NR sample. The main reason for NSs binding to the wrong positions on the DNA structure can be found in the NR purification process. Purifying NRs from NSs, the latter being a side product of our NR synthesis, leaves 12 ±3 % of NSs among the NRs, as deducted from TEM analysis of the sample NR– –NR. Consequently, also the proportion of NS in the sample NR–NS–NR is increased to 41 ±3 %, which is reflected in a higher extinction spectrum at the NS resonance frequency range for both samples.

Interestingly, slight variations in the positioning of the rods in our simulations resulted in strikingly different strengths of the calculated CD responses (Fig. 4). The general behavior can be understood best when thinking about the extreme arrangement of the two nanorods crossing each other in their midpoints. In this case the structure would depict an achiral "+" and hence no CD would be observable. In our simulations we moved the NRs outwards, starting from a structure lying between a + and an L shape (NRP 1). Consequently, only a weak CD response is generated. However, if the ends of the nanorods were moved outwards (NRP 2-7), the CD response of the longitudinal modes as well as the CD signal around the nanosphere resonance wavelength grew steadily stronger. We hypothesize that such a strong dependency can be a result of the almost matching resonance wavelength between the nanospheres (530 nm) and the transversal plasmon mode of the nanorods (516 nm).

To summarize, we studied the effect of a spherical transmitter NP in a chiral structure between a NR pair in experiment and simulation. Using the DNA origami technique, we obtained excellent control over the sample geometry and were able to assemble chiral transfer structures with high yields and precision. The chirality response of the NRs can be coupled strongly to a spherical plasmonic particle residing between the rods where it acts as transmitter. Our matching computational simulations corroborate our understanding of the system and all features of the CD signal can be explained specifically through coupling via the hotspots and generally through the near-field around the plasmonic particles. The synthesized nanostructures could potentially act as a new type of plasmonic chiral sensors or as transmitter elements for chiral spin locking in optical circuits.

**Supplementary Information** containing additional data, protocols and details on calculations is linked to the online version of the paper.

**Acknowledgements:** Kevin Martens and Tim Liedl are grateful for financial support through the Deutsche Forschungsgemeinschaft (DFG, German Research Foundation) through the SFB1032 (Project A6) and the ERC Consolidator Grant 818635 "DNA Funs". Felix Binkowski and Sven Burger acknowledge funding by the Deutsche Forschungsgemeinschaft under Germany´s Excellence Strategy – The Berlin Mathematics Research Center MATH+ (EXC-2046/1, project ID: 390685689) and by the German Federal Ministry of Education and Research (BMBF Forschungscampus MODAL, project number 05M20ZBM).

**Author Contributions:** KM, AOG and TL designed the research, KM and TL designed the nanostructures, KM produced the structures and performed the experiments. LN synthesized the nanoparticles. FB, LH, AOG and SB performed theoretical calculations. All authors wrote the manuscript.

**Author Information**: Reprints and permissions information is available. The authors declare no competing financial interests. Correspondence and requests for materials should be addressed to TL (tim.liedl@lmu.de).

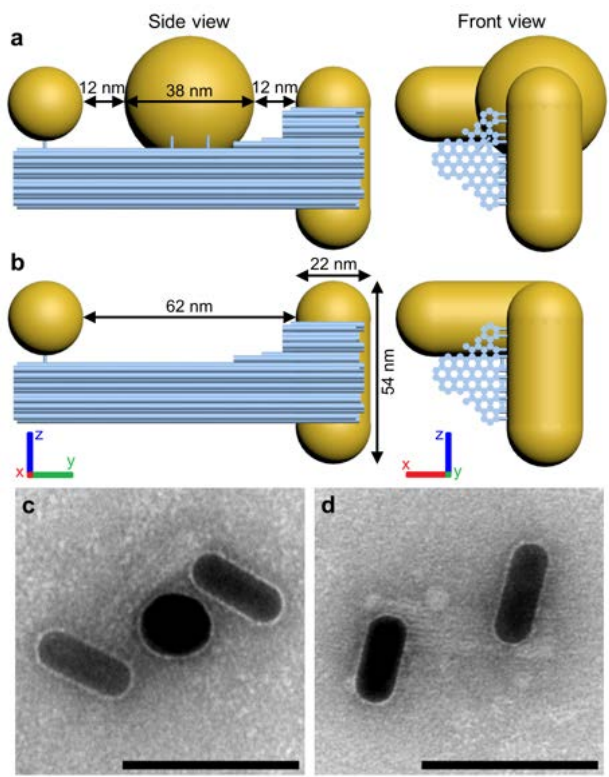

**Figure 1 | Chiral plasmonic transmitter. a**, Side view and front view of DNA origami-nanoparticle assemblies in a nanorod–nanosphere–nanorod (NR–NS–NR) arrangement and **b,** a nanorod–void–nanorod (NR– –NR) arrangement. The nanorods and the nanosphere are mounted on a DNA origami structure (blue cylinders represent DNA helices) via thiolated DNA strands that are anchored to the origami structure. **c,** Transmission electron micrograph of assemblies in the NR–NS–NR arrangement and **d**, in the NR– –NR arrangement. Scale bars: 100 nm

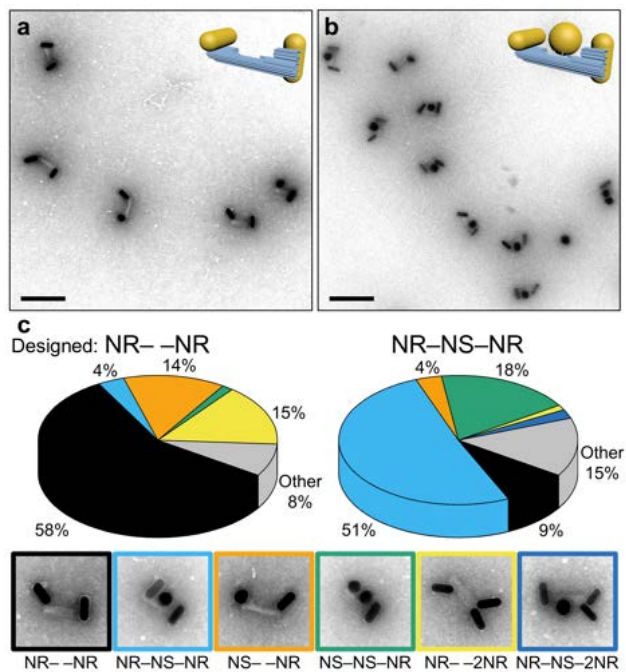

**Figure 2 | Assembly yields. a,** Electron micrograph of sample NR– –NR, and **b**, of sample NR–NS–NR. Scale bar: 200 nm. **c**, Distribution of assemblies in the samples NR– –NR and NR–NS–NR. Structures assembled as designed pose the majority with a variety of misassemblies occurring. Approximately 500 individual assemblies were studied (cf. Supplementary Note S3).

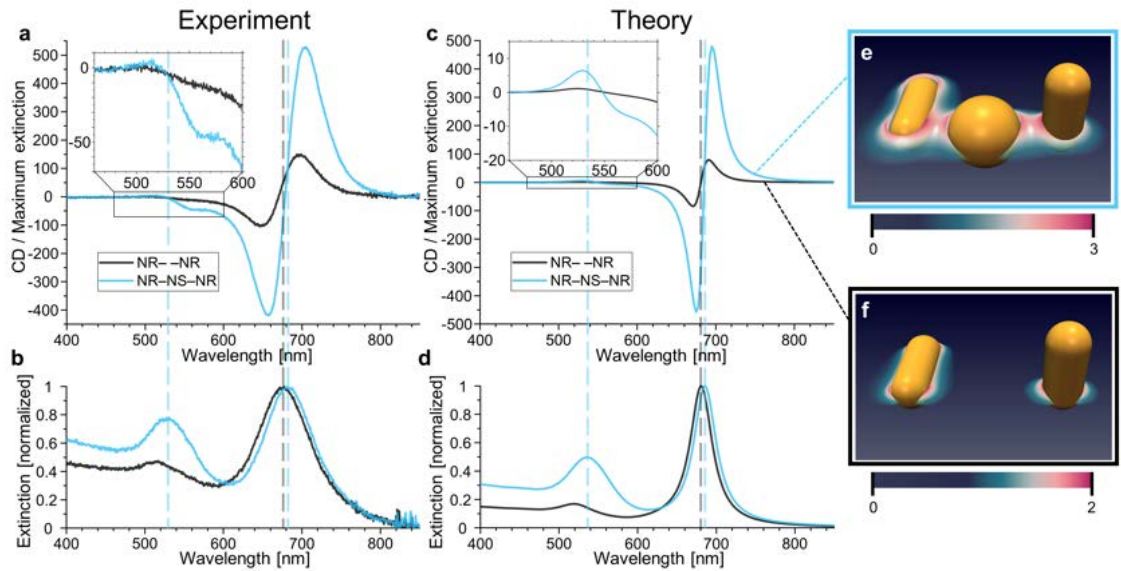

**Figure 3 | CD transfer experiment and theory. a**, CD and **b**, extinction measurements of samples NR– –NR and NR–NS–NR, normalized by the maximum extinction value. **c**, Simulated CD signal and **d**, simulated extinction of samples NR– –NR and NR–NS–NR. The signal strength was matched to the experimental amplitude for the CD. **e**, Electric near-field intensity around the plasmonic particles shown as a heat map for the arrangement NR–NS–NR and **f**, for the arrangement NR– –NR.

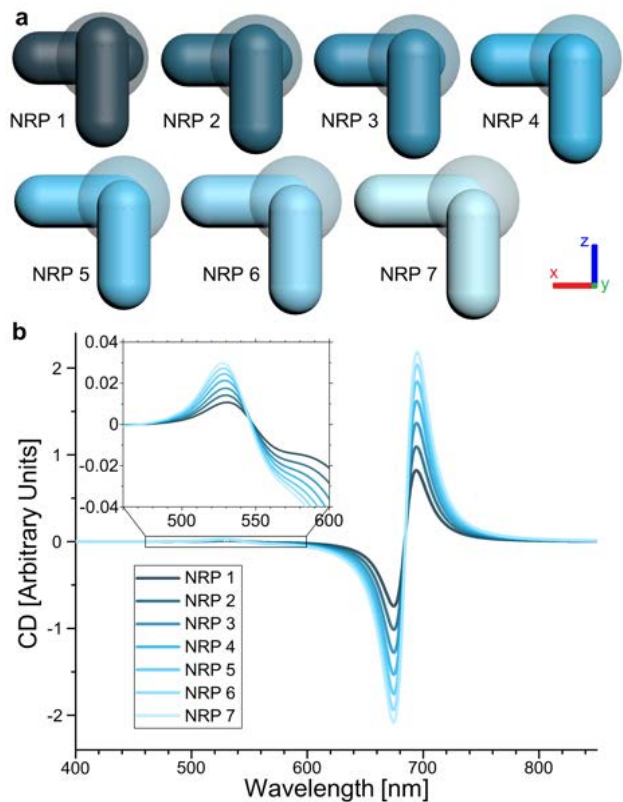

**Figure 4 | CD intensity depends critically on nanoparticle position. a**, Slightly varying nanorod positions (NRP) are exemplified for the NR–NS–NR assemblies. In each step, the lower NR is shifted 2 nm in x-direction and the upper NR is shifted 2 nm in negative z-direction. **b**, Simulated CD signals of NR–NS–NR assemblies in NRP 1-7, starting with a weak signal for NRP 1 and ending with a much stronger signal for NRP 7.

# Supplementary Information

# Long- and Short-Ranged Chiral Interactions in DNA-assembled Plasmonic Chains

*Kevin Martens[1], Felix Binkowski[2], Linh Nguyen[1], Li Hu[3], Alexander O. Govorov[4], Sven Burger[2,5] and Tim Liedl[1*]*

[1]*Faculty of Physics, Ludwig-Maximilians-University, Geschwister-Scholl-Platz 1, D-80539 Munich, Germany*

[2]*Zuse Institute Berlin, Takustraße 7, D-14195 Berlin, Germany*

[3]*Chongqing Engineering Laboratory for Detection, Control and Integrated System, Chongqing Technology and Business University, Chongqing 400067, China*

[4]*Department of Physics and Astronomy, Nanoscale and Quantum Phenomena Institute, Ohio University, Athens, Ohio 45701, United States*

[5]*JCMwave GmbH, Bolivarallee 22, D-14050 Berlin, Germany*

\* tim.liedl@lmu.de

## Materials

DNA scaffold strands (p8064) were prepared following previously described procedures.[1,2] Unmodified staple strands (purification: desalting) were purchased from Eurofins MWG. Thiol-modified strands (purification: HPLC) were purchased from Biomers. Uranyl formate for negative TEM staining was purchased from Polysciences, Inc.. Spherical gold nanoparticles were purchased from BBI Solutions. Other chemicals were purchased from CarlRoth and Sigma-Aldrich.

# Supplementary Note S1: DNA Origami Design

The left half of the dimeric origami structure (yellow part in Figure S1) was folded with an 8064 base pair (bp) scaffold strand and 126 core staple strands as well as 62 "$C_4$ endcap" staples, 22 staples for dimerization and 12 "handle" strands for NP assembly. The right half of the origami structure (blue part in Figure S1) was folded separately but with the same 8064 bp scaffold strand and a different set of 123 core staple strands as well as 70 $C_4$ endcap staples, 25 staples for dimerization and 11 handle strands for NP assembly.

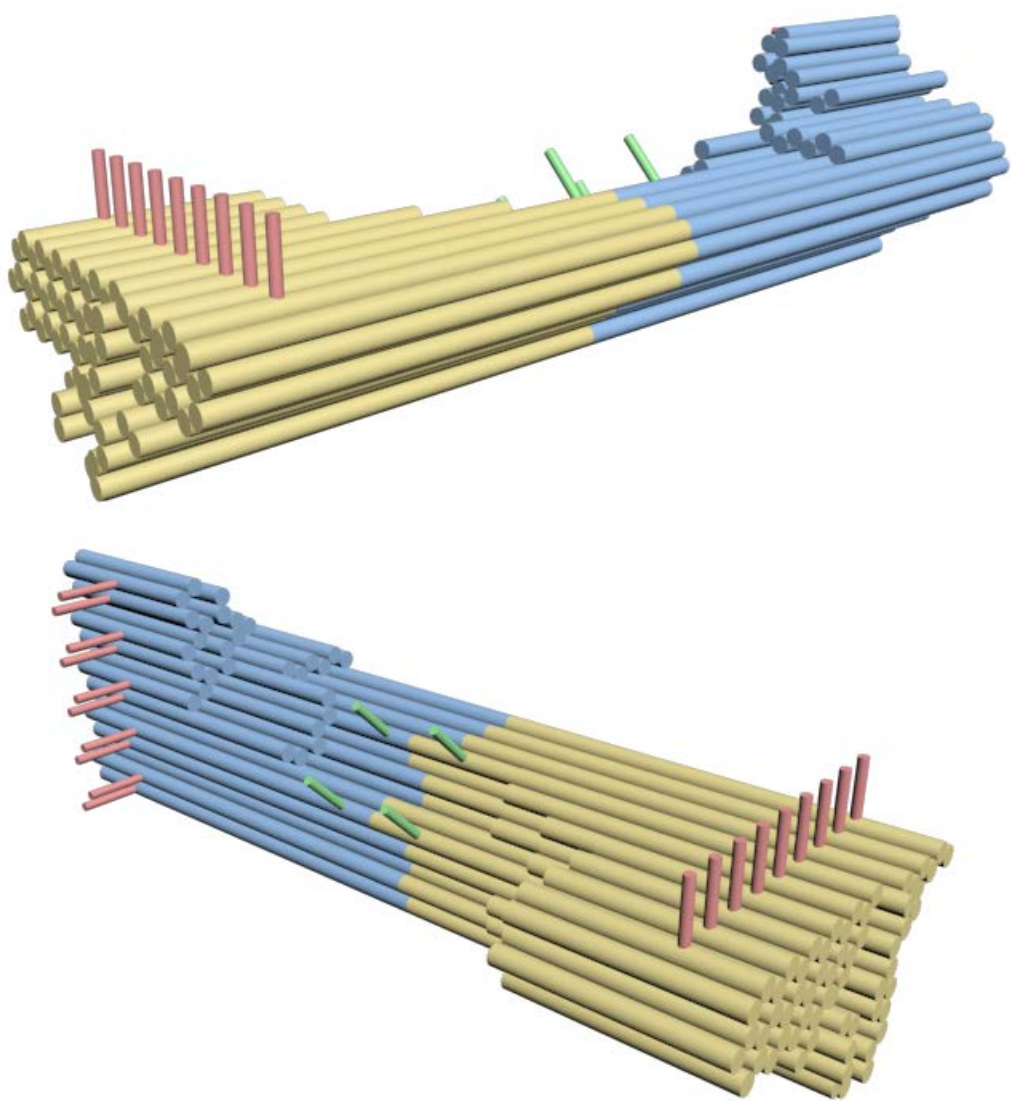

**Figure S1:** 3D Model of the DNA origami structure, depicting the individual halfs in blue and yellow (cylinders represent DNA helices). The handles for the NS are depicted in green as well as the handles for the NRs in red.

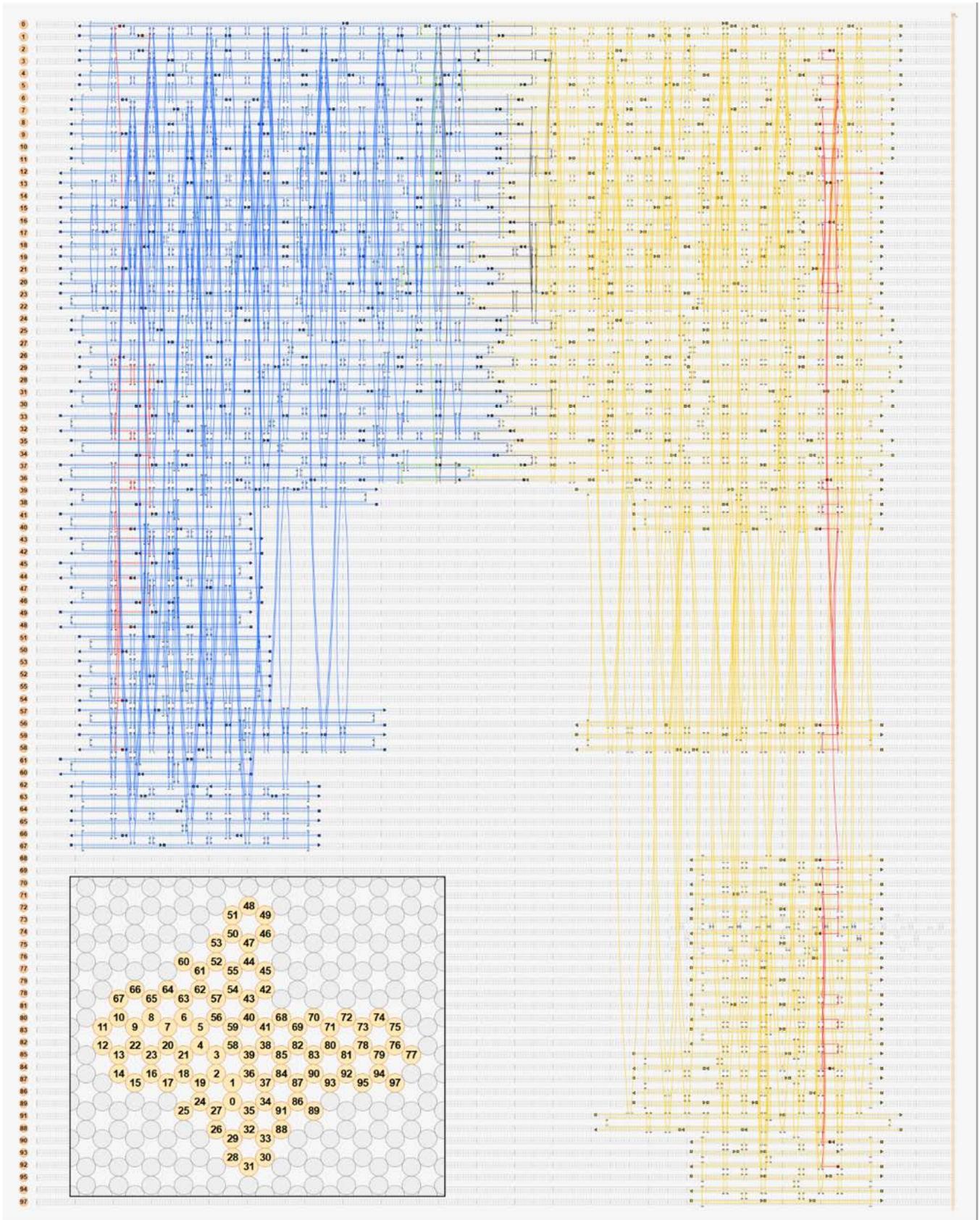

**Figure S2:** Cadnano design of the DNA origami structure, depicting the individual halves in blue and yellow as well as the handles for the NS in green and the handles for the NRs in red. The staples that connect the individual sides are depicted in black.

# Supplementary Note S2: DNA Origami Synthesis

The scaffold was used during folding at a concentration of 16 nM with 160 nM core and endcap staples plus 200 nM dimer and handle staples. 1X TE and 28 mM $MgCl_2$ were added as buffer solution. The two solutions containing the two mixtures were heated up to 65°C and then cooled down to room temperature over the course of 24 h. Subsequently the left side of the origami and the right side were combined in equal amounts and left to dimerize over 48 h. Dimer origamis were purified using gel electrophoresis with 0.7% agarose gel in a buffer of 1X TAE, 11 mM $MgCl_2$ and 0.05‰ Roti Stain as intercalating dye. The gel was run for 2.5 h at 70 V, before the origami dimer band was cut out under UV light and afterwards squeezed to redisperse the sample in buffer.

Table S1: Left half DNA origami protocol.

| Component | Concentration | Amount | End Concentration |
| --- | --- | --- | --- |
| Scaffold 8064 | 100 nM | 16 µL | 16nM |
| Core Staples | 397 nM | 40.3 µL | 160nM |
| Endcap Staples | 806 nM | 19.9 µL | 160nM |
| Dimere Staples | 2273 nM | 8.8 µL | 200nM |
| Handles | 4167 nM | 4.8 µL | 200nM |
| TE | 20X | 5 µL | 1X |
| MgCl2 | 1 M | 2.8 µL | 28mM |
| H2O | - | 2.4 µL | - |
| **Total** | **100 nM** | **100 µL** | |

Table S2: Right half DNA origami protocol.

| Component | Concentration | Amount | End Concentration |
|---|---|---|---|
| Scaffold 8064 | 100 nM | 16 µL | 16nM |
| Core Staples | 407 nM | 39.3 µL | 160nM |
| Endcap Staples | 714 nM | 22.4 µL | 160nM |
| Dimere Staples | 2000 nM | 10 µL | 200nM |
| Handles | 4545 nM | 4.4 µL | 200nM |
| TE | 20X | 5 µL | 1X |
| MgCl2 | 1 M | 2.8 µL | 28mM |
| H2O | - | 0.1 µL | - |
| **Total** | **100 nM** | **100 µL** | |

For TEM analysis, samples were incubated for 15 min on copper grids (Ted Pella Inc., Redding, USA) before being dabbed off with a filter paper and subsequently stained with 2% uranyl format in two steps. In the first step the uranyl format solution only quickly washes the grid, in the second step it is left to incubate for 15 s before being dabbed off. Images were taken with a JEOL JEM 1011 electron microscope at 80 kV.

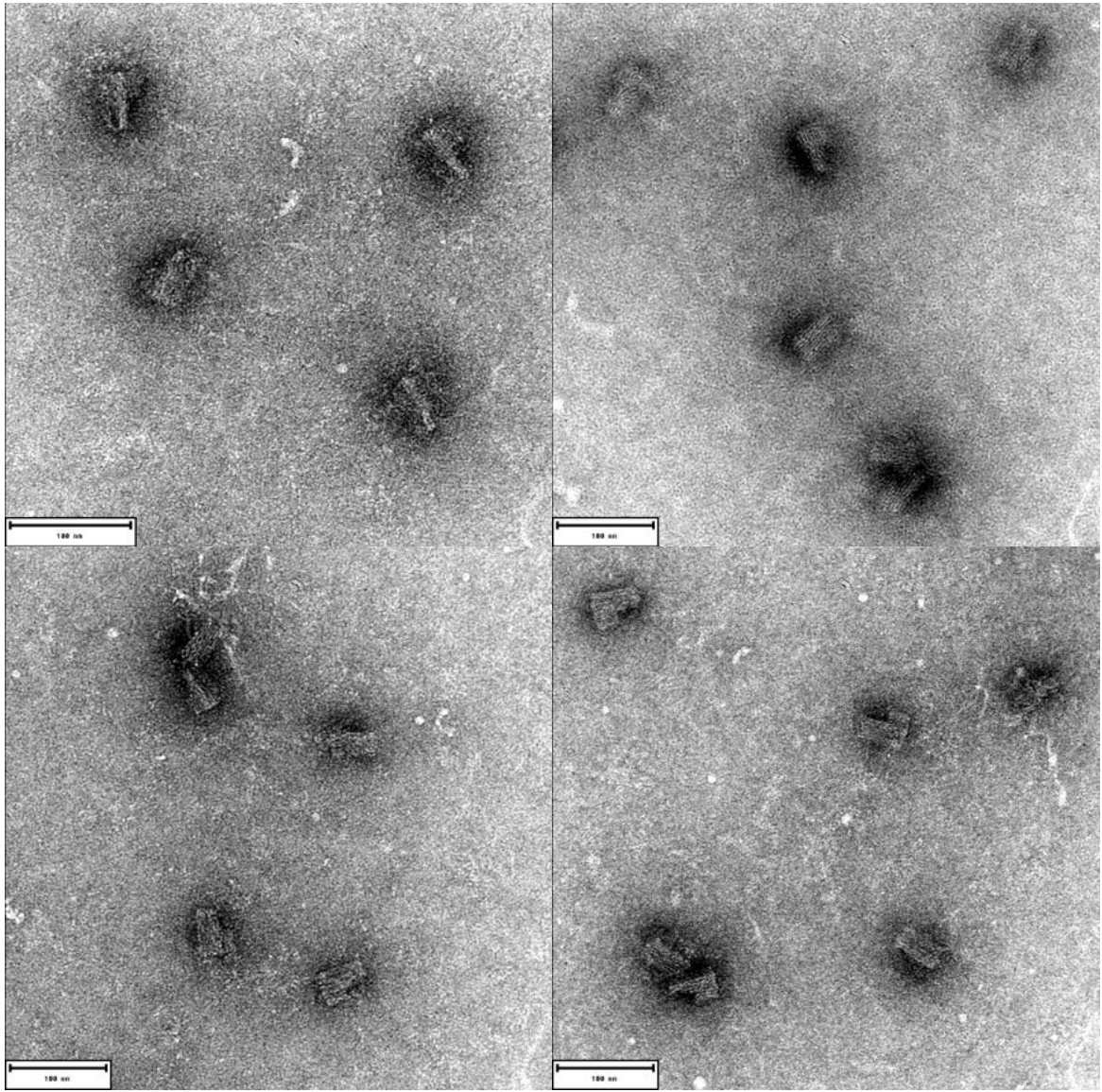

**Figure S3:** Electron micrographs of the left half of the DNA origami after gel electrophoresis purification

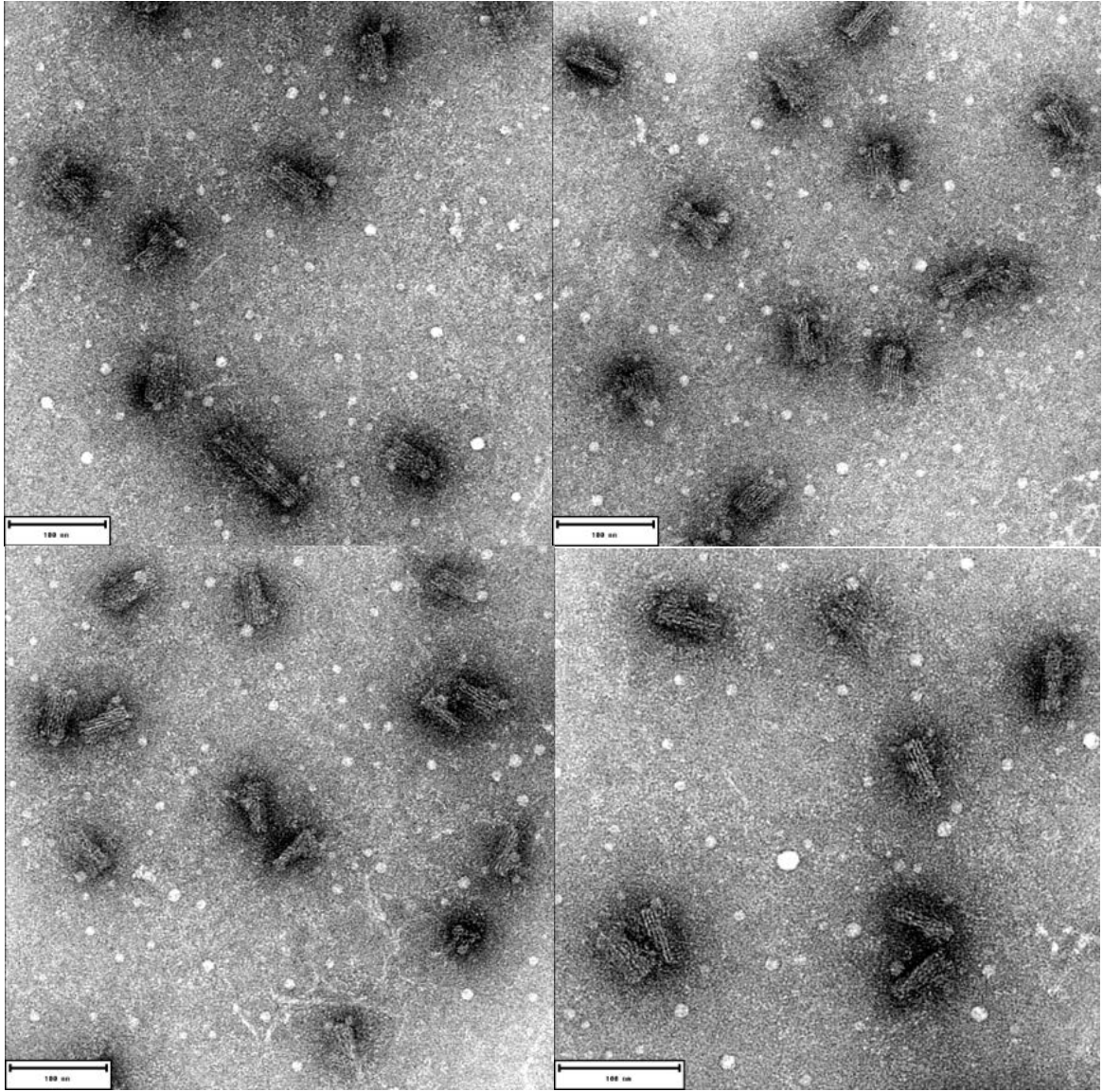

**Figure S4:** Electron micrographs of the right half of the DNA origami after gel electrophoresis purification

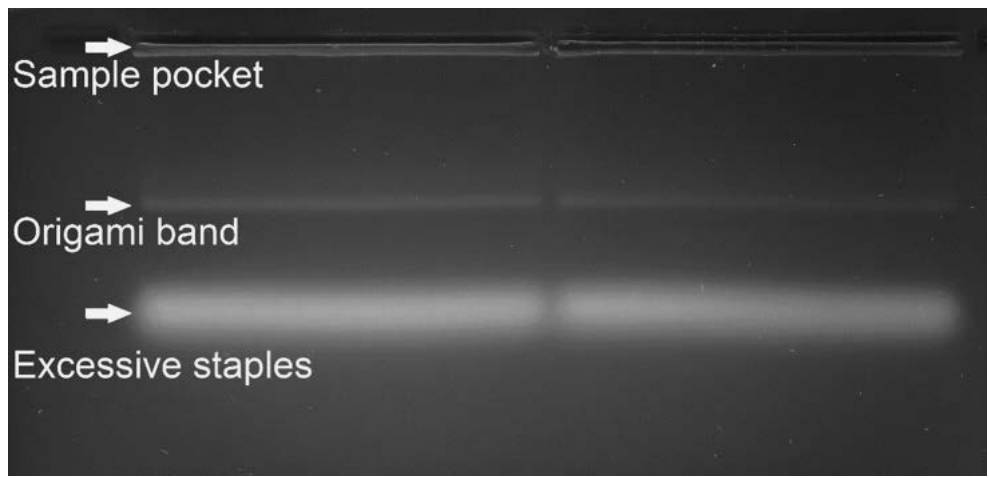

**Figure S5:** Gel electrophoresis band of DNA origami structure after dimerization

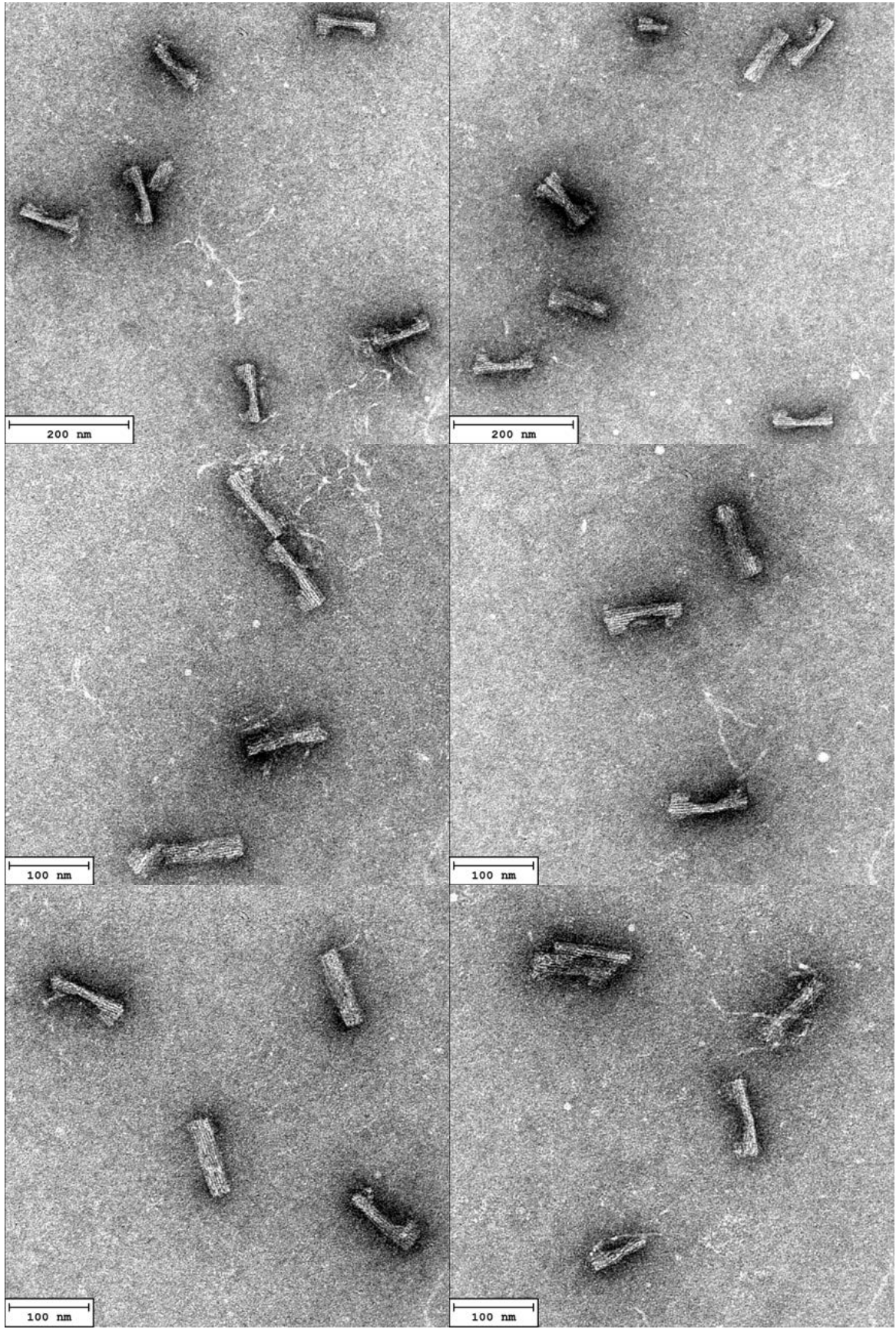

**Figure S6:** Electron micrographs of the complete DNA origami structure (purified)

# Supplementary Note S3: DNA Origami-Nanoparticles Assembly Synthesis

40 nm NSs were incubated at an optical density (OD) of 4 with 10 mM thiol-modified DNA oligonucleotides, previously activated with TCEP, and 0.02% SDS. NR synthesis was performed following the protocol of Ye et al.[3] 60 nm x 23 nm NRs were incubated at OD 1.4 with 5 mM thiol-modified DNA oligonucleotides and 0.1% SDS. Samples were frozen, thawed and purified using gel electrophoresis with a 0.7% agarose gel in a buffer of 1X TAE, 11 mM $MgCl_2$, run for 1.5 h at 120 V. Subsequently the correct monomer bands were cut and squeezed to redisperse in buffer.

For the synthesis of the NR– –NR sample, NRs were added to the origami structures in a ratio of 10:1 in a buffer of 1X TAE, 11 mM MgCl2 plus 500 mM NaCl and incubated for 24 h. The sample was purified using gel electrophoresis with a 0.7% agarose gel in a buffer of 1X TAE, 11 mM $MgCl_2$, run for 1.5 h at 70 V. The band of the structures was cut out and squeezed. For the NR–NS–NR sample, first NSs were incubated with the DNA origami in a ratio of 5:1 in a buffer of 1X TAE, 11 mM $MgCl_2$ plus 500 mM NaCl for 24 h. Afterwards NRs were added in a ratio of 10:1 to the origami, and incubated in the same buffer for 24 h. The samples were purified using gel electrophoresis with a 0.7% agarose gel in a buffer of 1X TAE, 11 mM $MgCl_2$, run for 1.5 h at 70 V. The structure bands were cut out, squeezed and identified by TEM.

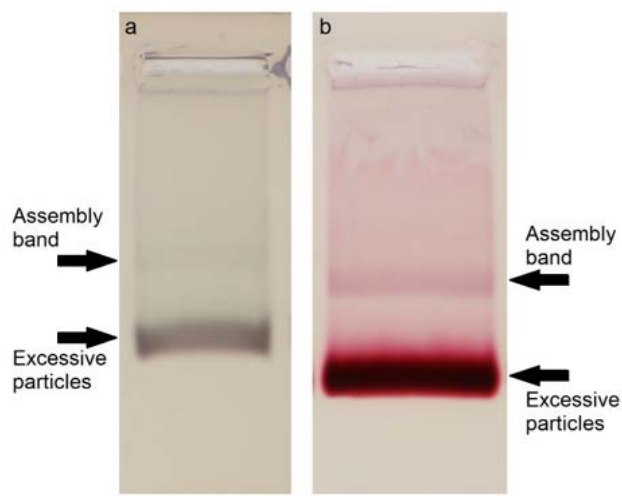

**Figure S7:** Gel electrophoresis bands of samples with a) the NR– –NR arrangement and (b) the NR–NS–NR arrangement

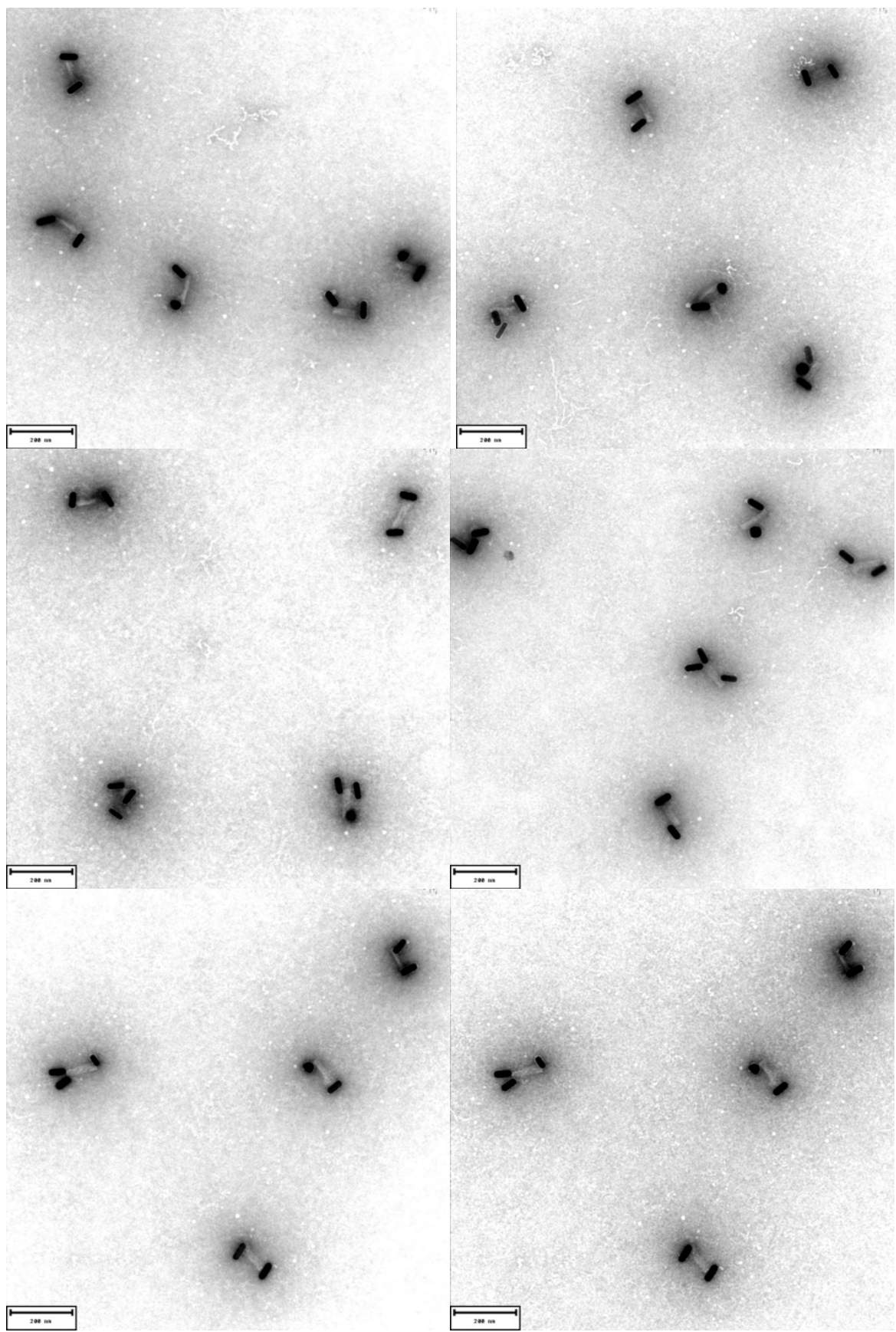

**Figure S8:** Electron micrographs of the NR– –NR arrangement

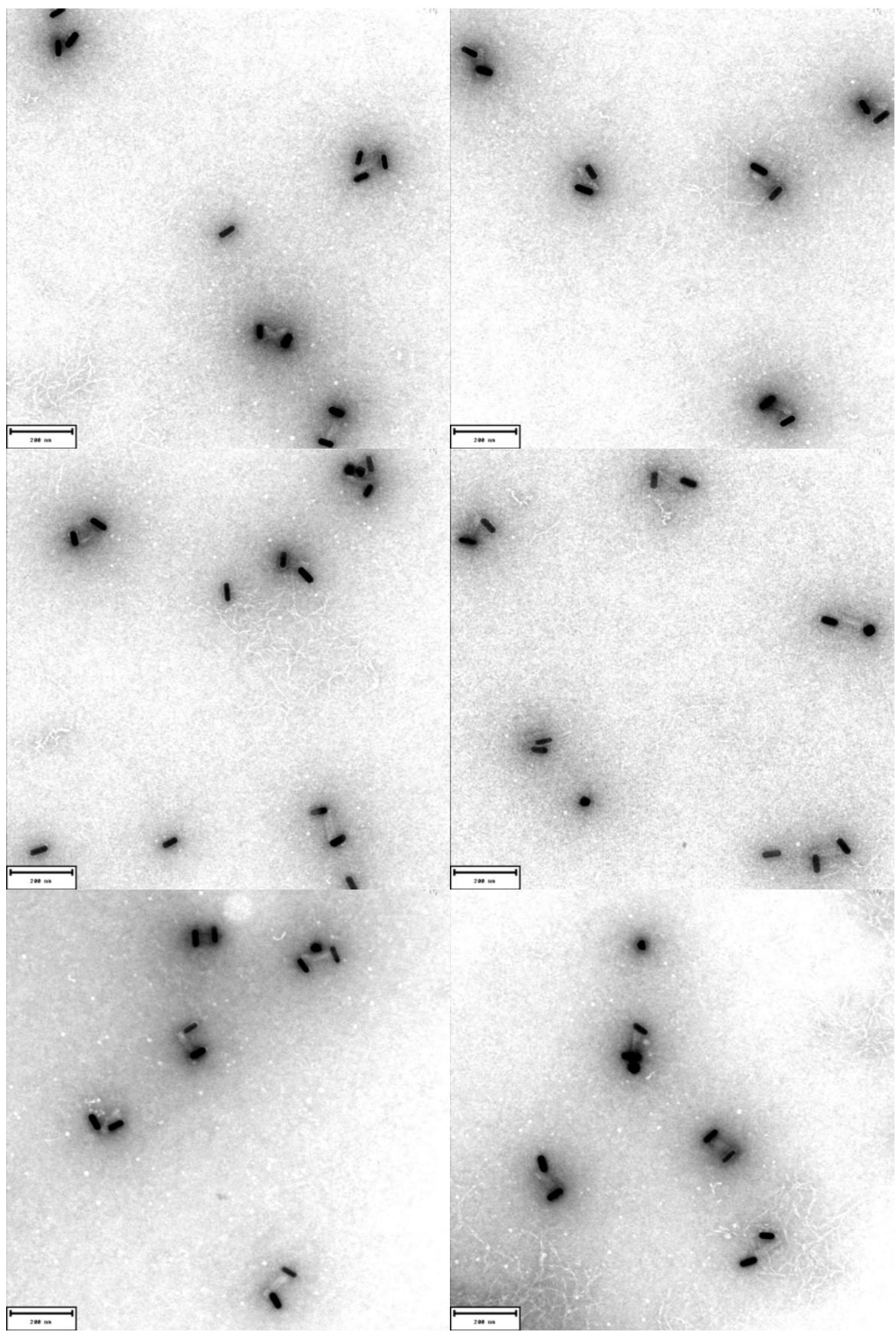

**Figure S9:** Electron micrographs of the NR– –NR arrangement

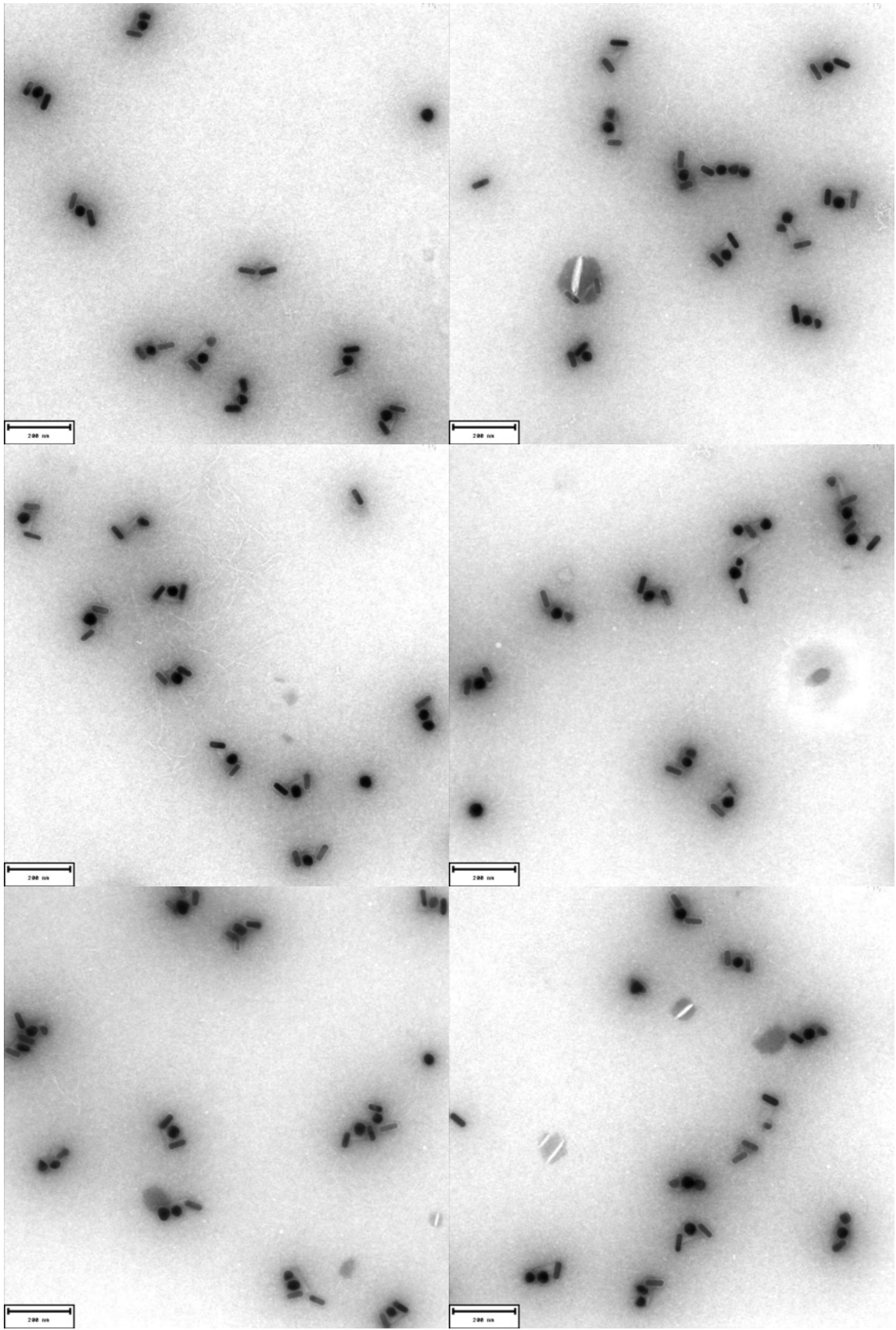

**Figure S10:** Electron micrographs of NR–NS–NR sample

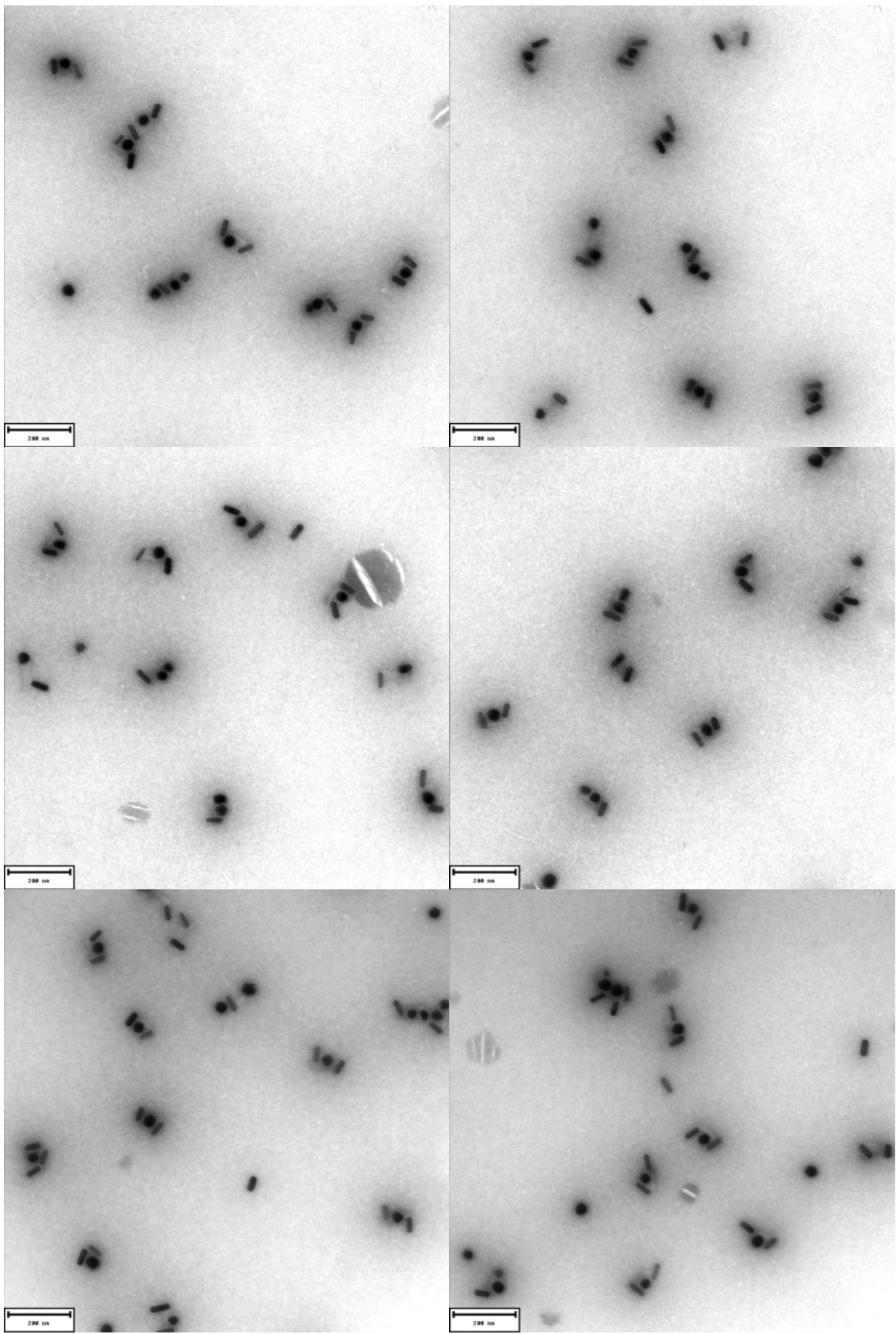

**Figure S11**: Electron micrographs of NR–NS–NR sample

**Table S3:** NR– –NR Synthesis Assembly Statistics.

| Assembly | Number | Percentage |
| --- | --- | --- |
| NR– – | 3 | < 3 % |
| NR– –NR | 113 | 57.5 ±7 % |
| NS– –NR | 27 | 14 ±4.5 % |
| NS– –NS | 1 | < 1.5 % |
| NR– –2NR | 30 | 15.5 ±5 % |
| NR–NS–NR | 7 | 3.5 ±2.5 % |
| NR–NR–NS | 8 | 4 ±2.5 % |
| NR–NS–NS | 3 | < 3 % |
| NS–NR–NS | 1 | < 1.5 % |
| NR–NR–NR–NR | 2 | < 2 % |
| NR–NR–NS–NS | 1 | < 1.5 % |
| **Total** | **196** | **100 %** |

**Table S4:** NR– –NR Synthesis Particle Statistics.

| Particle | Number | Percentage |
| --- | --- | --- |
| NS | 54 | 12 ±3 % |
| NR | 390 | 88 ±3 % |
| **Total** | **444** | **100 %** |

**Table S5:** NR–NS–NR Synthesis Assembly Statistics.

| Assembly | Number | Percentage |
| --- | --- | --- |
| NR– –NR | 28 | 9 ±3 % |
| NS– –NR | 11 | 3.5 ±2 % |
| NS– –NS | 3 | < 2 % |
| NR– –2NR | 4 | 1.5 ±1 % |
| NR–NS–NR | 160 | 51 ±5.5 % |
| NS– –2NR | 8 | 2.5 ±1.5 % |
| NR–NS–NS | 57 | 18 ±4.5 % |
| NS–NR–NS | 4 | 1.5 ±1 % |
| NS–NS–NS | 5 | 1.5 ±1.5 % |
| NR–2NS–NR | 5 | 1.5 ±1.5 % |
| NR–NR–2NR | 3 | < 2 % |
| NS–NS–2NR | 3 | < 2 % |
| NR–NS–NR–NS | 7 | 2 ±1.5 % |
| NR–NS–2NR | 7 | 2 ±1.5 % |
| NR–NS–2NS | 4 | 1.5 ±1 % |
| NS–NR–2NS | 3 | < 2 % |
| NS–NR–3NS | 1 | < 0.5 % |
| **Total** | **313** | **100 %** |

**Table S6:** NR– –NR Synthesis Particle Statistics.

| Particle | Number | Percentage |
|---|---|---|
| NS | 354 | 41 ±3 % |
| NR | 577 | 59 ±3 % |
| **Total** | **931** | **100 %** |

## Supplementary Note S4: CD and Extinction Measurements

Samples were measured with a Chirascan circular dichroism spectrometer (Applied Photophysics, Surrey, UK) in cuvettes with 3 mm pathlengths. Spectra were collected in 0.5 nm steps with 0.3 s for each step. 3 measurements were made an averaged for the NR– –NR sample.

## Supplementary Note S5: Numerical Simulations

For solving Maxwell's equations we use a higher-order finite-element method (FEM), implemented in the solver JCMsuite.[4] The geometry is discretized using a tetrahedral mesh with curvilinear mesh elements along the curved surfaces of the NPs. Transparent boundary conditions are realized by using perfectly matched layers. We use tabulated material data for NP material Au[40] and a constant refractive index of 1.4 for the background material. High numerical accuracy is ensured by using a conservative setting of the numerical parameters (polynomial degree of the FEM ansatz functions of $p=2$ and mesh element edge size smaller than 7 nm for the NPs and 14 nm for background material). The absorbed electromagnetic field energy and the electromagnetic field energy scattered outwards, corresponding to each circular polarized source term (left-hand circular polarization, LCP, and right-hand circular polarization, RCP), are obtained in post-processes. The extinction is given by the sum of absorption and scattering for both polarization directions, LCP and RCP, and for all six directions of incidence. The CD (g-factor) is given by the difference between absorption and scattering for LCP and absorption and scattering for RCP, and is normalized to the extinction maximum of the respective wavelength spectrum. Field patterns for visualization purposes (Fig. 3e, f) are obtained by exporting the computed near-fields on specific cross-sections, and a summation of the exported fields over all source terms at a specific wavelength. For performing numerical parameter studies as shown in Fig. 4, the physical quantities of the project are parameterized, and a scripting language (Matlab) is used to automatically generate the input files and to distribute the FEM computations to various threads on a workstation for parallel computation of the parameter- and wavelength-scans.